\documentclass[reprint,superscriptaddress,aps]{revtex4-1}
\usepackage{amsmath}
\usepackage{amssymb}
\usepackage{bm}
\usepackage{braket}
\usepackage{color}
\usepackage[varg]{txfonts}
\usepackage{varwidth}
\usepackage{dcolumn}
\usepackage[breaklinks,colorlinks=true,linkcolor=blue,urlcolor=cyan,citecolor=blue]{hyperref}
\usepackage{here}
\usepackage{graphicx}

\begin{document}

\title{Breathing pyrochlore magnet CuGaCr$_{4}$S$_{8}$: Magnetic, thermodynamic, and dielectric properties}

\author{Masaki Gen}
\email{masaki.gen@riken.jp}
\affiliation{Department of Advanced Materials Science, The University of Tokyo, Kashiwa 277-8561, Japan}
\affiliation{RIKEN Center for Emergent Matter Science (CEMS), Wako 351-0198, Japan}

\author{Hajime Ishikawa}
\affiliation{Institute for Solid State Physics, The University of Tokyo, Kashiwa 277-8581, Japan}

\author{Atsushi Miyake}
\affiliation{Institute for Solid State Physics, The University of Tokyo, Kashiwa 277-8581, Japan}

\author{Takeshi Yajima}
\affiliation{Institute for Solid State Physics, The University of Tokyo, Kashiwa 277-8581, Japan}

\author{Harald O. Jeschke}
\affiliation{Research Institute for Interdisciplinary Science, Okayama University, Okayama 700-8530, Japan}

\author{Hajime Sagayama}
\affiliation{Institute of Materials Structure Science, High Energy Accelerator Research Organization, Tsukuba 305-0801, Japan}

\author{Akihiko Ikeda}
\affiliation{Institute for Solid State Physics, The University of Tokyo, Kashiwa 277-8581, Japan}
\affiliation{Department of Engineering Science, University of Electro-Communications, Chofu, Tokyo 182-8585, Japan}

\author{Yasuhiro H. Matsuda}
\affiliation{Institute for Solid State Physics, The University of Tokyo, Kashiwa 277-8581, Japan}

\author{Koichi Kindo}
\affiliation{Institute for Solid State Physics, The University of Tokyo, Kashiwa 277-8581, Japan}

\author{Masashi Tokunaga}
\affiliation{Institute for Solid State Physics, The University of Tokyo, Kashiwa 277-8581, Japan}

\author{Yoshimitsu Kohama}
\affiliation{Institute for Solid State Physics, The University of Tokyo, Kashiwa 277-8581, Japan}

\author{Takashi Kurumaji}
\affiliation{Department of Advanced Materials Science, The University of Tokyo, Kashiwa 277-8561, Japan}

\author{Yusuke Tokunaga}
\affiliation{Department of Advanced Materials Science, The University of Tokyo, Kashiwa 277-8561, Japan}

\author{Taka-hisa Arima}
\affiliation{Department of Advanced Materials Science, The University of Tokyo, Kashiwa 277-8561, Japan}
\affiliation{RIKEN Center for Emergent Matter Science (CEMS), Wako 351-0198, Japan}

\begin{abstract}

We investigate the crystallographic and magnetic properties of a chromium-based thiospinel CuGaCr$_{4}$S$_{8}$.
From a synchrotron x-ray diffraction experiment and structural refinement, Cu and Ga atoms are found to occupy the tetrahedral {\it A}-sites in an alternate way, yielding breathing pyrochlore Cr network.
CuGaCr$_{4}$S$_{8}$ undergoes a magnetic transition associated with a structural distortion at 31~K in zero magnetic field, indicating that the spin-lattice coupling is responsible for relieving the geometrical frustration.
When applying a pulsed high magnetic field, a sharp metamagnetic transition takes place at 40~T, followed by a 1/2-magnetization plateau up to 103~T. 
These phase transitions accompany dielectric anomalies, suggesting the presence of helical spin correlations in low-field phases.
The density-functional-theory calculations suggest that CuGaCr$_{4}$S$_{8}$ is dominated by antiferromagnetic and ferromagnetic exchange couplings within small and large tetrahedra, respectively, in analogy with CuInCr$_{4}$S$_{8}$.
We argue that {\it A}-site-ordered Cr thiospinels serve as an excellent platform to explore diverse magnetic phases along with pronounced magnetoelastic and magnetodielectric responses.

\end{abstract}

\date{\today}
\maketitle

\section{\label{Sec1}Introduction}

\begin{figure}[t]
\centering
\includegraphics[width=\linewidth]{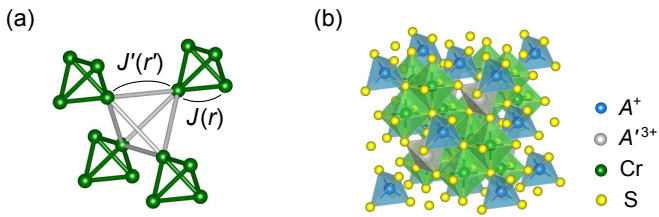}
\caption{(a) Schematic of a breathing pyrochlore lattice, where small and large tetrahedra with the nearest-neighbor bond lengths, $r$ and $r'$, are characterized by the exchange interactions, $J$ and $J'$, respectively. (b) Crystal structure of the {\it A}-site-ordered Cr spinels. The present focus is CuGaCr$_{4}$S$_{8}$, where nonmagnetic Cu$^{+}$ and Ga$^{3+}$ ions are found to be arranged in the zinc-blende-type structure, yielding a breathing pyrochlore Cr network. The illustrations are drawn with VESTA software \cite{2011_Mom}.}
\label{Fig1}
\end{figure}

Magnetic materials of the pyrochlore lattice, a three-dimensional network of corner-sharing tetrahedra, have been a central research subject in the context of frustrated magnetism \cite{2010_Gar, 2021_Rei}.
In recent years, the {\it breathing} pyrochlore lattice, where up- and down-pointing tetrahedra differ in size, has attracted growing interest from the viewpoint of ground state control \cite{2013_Oka}.
The key concept of this spin model lies in the introduction of inequivalent exchange couplings $J$ and $J'$ in the small and large tetrahedra, respectively [Fig.~\ref{Fig1}(a)].
Depending on the signs and magnitudes of $J$ and $J'$ as well as the nature of spins, various exotic magnetic states and emergent phenomena have been theoretically predicted: e.g., unconventional spin-liquid states and excitations \cite{2015_Ben, 2016_Sav, 2016_Li, 2019_Iqb, 2020_Yan, 2022_Han}, spin-lattice-coupled superlattice long-range orders (LROs) \cite{2019_Aoy, 2021_Aoy}, and a magnetic hedgehog-lattice \cite{2021_Aoy_HG, 2022_Aoy_HG}.

A representative realization of the breathing pyrochlore system is a quantum Heisenberg antiferromagnet Ba$_{3}$Yb$_{2}$Zn$_{5}$O$_{11}$, whose magnetism is governed by Yb$^{3+}$ ions with pseudospin-1/2 \cite{2014_Kim}.
In this compound, the breathing ratio $r'/r$, where $r$ ($r'$) represents the nearest-neighbor (NN) bond length in small (large) tetrahedra, amounts to approximately 2, resulting in $J \gg J' > 0$, i.e., close to the decoupled tetrahedron limit \cite{2014_Kim}.
The magnetization, specific heat, and inelastic neutron scattering suggested the singlet formation without any signs of a magnetic LRO at low temperatures \cite{2014_Kim, 2016_Hak}, though the development of intertetrahedral correlations is also pointed out \cite{2016_Rau, 2022_Dis}.

For the larger spin case, {\it A}-site-ordered chromium-based spinels {\it AA'}Cr$_{4}${\it X}$_{8}$, where Cr$^{3+}$ ions with spin-3/2 form a breathing pyrochlore lattice, offer a fertile playground to address an effective spin Hamiltonian with various sets of $J$ and $J'$ \cite{2019_Gho}.
Due to the difference in the ionic radius between {\it A}$^{+}$ and {\it A'}$^{3+}$ cations, their crystallographic ordering like the zinc-blende-type arrangement should modulate the chemical pressure and as a consequence induce the breathing bond-alternation in the Cr pyrochlore network [Fig.~\ref{Fig1}(b)].
This material design was first proposed by Joubert and Durif in 1966 \cite{1966_Jou}.
They prepared two oxides, LiGaCr$_{4}$O$_{8}$ and LiInCr$_{4}$O$_{8}$, and found the lack of an inversion center in their crystal structures, signaling the ordering of Li and Ga/In atoms.
Subsequently, Pinch {\it et al}. synthesized Cr thiospinels {\it AA'}Cr$_{4}$S$_{8}$ with various combinations of {\it A} and {\it A'} atoms: {\it A}=Li, Cu, Ag; {\it A'}=Al, Ga, In \cite{1970_Pin}.
Among them, the {\it A}-site ordering was confirmed for LiGaCr$_{4}$S$_{8}$, LiInCr$_{4}$S$_{8}$, CuAlCr$_{4}$S$_{8}$, and CuInCr$_{4}$S$_{8}$.
Structural refinements and detailed physical property measurements on these compounds have been actively performed in the last decade \cite{2015_Nil, 2016_Lee, 2016_Sah, 2018_Oka, 2018_Pok, 2019_Gen, 2020_Gen, 2020_Kan, 2020_Pok, 2021_Gao, 2022_Sha, 2022_Gen}, triggered by the renewed interest by Okamoto {\it et al.} in 2013 \cite{2013_Oka}.

It is noteworthy that a peculiar combination of antiferromagnetic (AFM) $J$ and ferromagnetic (FM) $J'$ can be realized in {\it AA'}Cr$_{4}${\it X}$_{8}$ due to the competition between the AFM Cr--Cr direct exchange and FM Cr--{\it X}--Cr superexchange interactions \cite{2019_Gho}.
Such a spin Hamiltonian can be effectively mapped on the spin-6 Heisenberg antiferromagnet on the face-centered-cubic (FCC) lattice if $J'$ is strong.
Indeed, a cluster excitation was observed in CuInCr$_{4}$S$_{8}$ at low temperatures \cite{2021_Gao}, indicating the development of FM correlations within each large tetrahedron.
Another characteristic feature is the intrinsic strong spin-lattice coupling (SLC) arising from the sensitivity of the strength of $J$ ($J'$) against the NN Cr--Cr bond length, i.e., large $|dJ/dr|$ ($|dJ'/dr'|$).
The SLC can act as a principal perturbation to lift the macroscopic degeneracy and bring about magnetostructural transitions at low temperatures \cite{2015_Nil, 2016_Lee, 2016_Sah, 2020_Kan} and in an applied magnetic field \cite{2019_Gen, 2020_Gen, 2021_Aoy}.
Interestingly, CuInCr$_{4}$S$_{8}$ exhibits a fascinating magnetic-field-versus-temperature ($H$-$T$) phase diagram including a robust 3-up-1-down phase associated with a 1/2-magnetization plateau \cite{2020_Gen} as well as a thermal equilibrium phase pocket \cite{2022_Gen} reminiscent of a skyrmion lattice \cite{2009_Muh, 2012_Sek, 2012_Oku}.

In this work, we report the structural, magnetic, thermodynamic, and dielectric properties of CuGaCr$_{4}$S$_{8}$.
Although CuGaCr$_{4}$S$_{8}$ was previously synthesized in Refs.~\cite{1970_Pin, 1976_Wil, 2005_Kes, 2018_Ami}, no conclusive remark on the {\it A}-site ordering was given because the close proximity of the scattering factors between Cu$^{+}$ and Ga$^{3+}$ made it challenging to judge the presence or absence of reflections forbidden for the {\it A}-site disordered case, i.e., $(hk0)$ peaks with $h+k=4n+2$.
Our synchrotron powder x-ray diffraction (XRD) measurement confirms the presence of 200 and 420 peaks of structural origin.
The Rietveld analysis shows that the {\it A}-site-ordered spinel structure with the breathing pyrochlore Cr network ($F{\overline 4}3m$ space group) provides better refinement than the normal spinel structure ($Fd{\overline 3}m$ space group).
CuGaCr$_{4}$S$_{8}$ undergoes a magnetic transition at $T_{\rm N} = 31$~K in zero magnetic field.
We observe a crystal symmetry lowering below $T_{\rm N}$, which is compatible with the reported low-symmetry incommensurate helical structure and was overlooked in the previous neutron diffraction study \cite{1976_Wil}.
A series of pulsed high-field experiments reveal that CuGaCr$_{4}$S$_{8}$ exhibits a rich $H$-$T$ phase diagram associated with magnetoelastic and magnetodielectric effects, similar to that of CuInCr$_{4}$S$_{8}$ \cite{2020_Gen, 2022_Gen}.
The effective spin Hamiltonian of CuGaCr$_{4}$S$_{8}$ is discussed on the basis of the density-functional-theory (DFT) energy mapping as well as magnetoelastic theory \cite{2020_Gen}.

\section{\label{Sec2}Methods}
Polycrystalline samples of CuGaCr$_{4}$S$_{8}$ were synthesized by the conventional solid-state reaction method.
Starting ingredients were high-purity gallium ingots (99.999\%) and copper (99.99\%), chromium (99.99\%), and sulfur (99.99\%) powders.
They were mixed in the stoichiometric ratio, sealed in an evacuated quartz tube, and heated at 400$^{\circ}$C for 24~h and then at 800$^{\circ}$C for 48~h in a box furnace.
Then, the sintering was repeated twice at 900$^{\circ}$C for 96~h after grinding and pelletizing the sintered products.

A powder synchrotron x-ray diffraction (XRD) profile was collected at room temperature on Photon Factory BL-8A.
The wavelength was $\lambda = 0.689739~\AA$.
The Rietveld analysis was performed using the RIETAN-FP program \cite{RIETAN}.
The temperature evolution of the powder XRD pattern was measured between 4 and 300~K using a commercial x-ray diffractometer (SmartLab, Rigaku) at the Institute for Solid State Physics (ISSP), University of Tokyo.
The incident x-ray beam was monochromatized by a Johansson-type monochromator with a Ge(111) crystal to select only Cu-$K\alpha$1 radiation.

\begin{figure}[t]
\centering
\includegraphics[width=\linewidth]{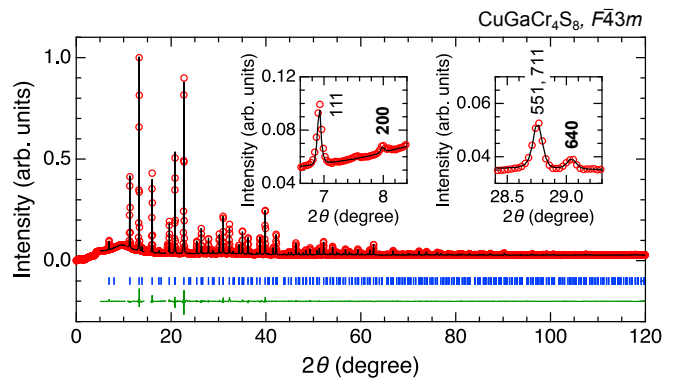}
\caption{Synchrotron XRD pattern of CuGaCr$_{4}$S$_{8}$ powder samples (red open circle) and calculated XRD pattern obtained by the Rietveld analysis (black solid line). The Rietveld refinement was performed in a range of $5^{\circ} < 2\theta < 120^{\circ}$, where several tiny peaks originating from Cr$_{3}$S$_{4}$ and unknown impurity phases are excluded. Blue vertical bars indicate the nuclear Bragg reflections, and the green line is the difference between the experimental and calculated patterns. Insets show enlarged views of the experimental XRD profiles focusing on 200 and 640 peaks, which are forbidden for $Fd{\overline 3}m$ but allowed for $F{\overline 4}3m$.}
\label{Fig2}
\end{figure}

Magnetization up to 7~T was measured using a SQUID magnetometer (MPMS, Quantum Design).
Magnetization up to 14~T was measured using a vibrating sample magnetometer installed in a physical property measurement system (PPMS, Quantum Design) equipped with a superconducting magnet.
Magnetization up to 57~T was measured by the induction method in a non-destructive (ND) pulsed magnet ($\sim$4~ms duration).
Magnetization up to 140~T was measured by the induction method using a coaxial-type pickup coil in a horizontal single-turn-coil (STC) megagauss generator ($\sim$8~$\mu$s duration) \cite{2020_Gen}.
Thermal expansion was measured by the fiber-Bragg-grating (FBG) method using an optical sensing instrument (Hyperion si155, LUNA) in a cryostat equipped with a superconducting magnet (Spectromag, Oxford) at zero field.
Longitudinal magnetostriction up to 54~T was measured by the FBG method in a ND pulsed magnet ($\sim$36 ms duration), where the optical filter method was employed to detect the relative sample-length change $\Delta L/L_{\rm 0T}$ \cite{2018_Ike}.
The fiber was attached to a rod-shaped sintered sample with epoxy Stycast1266.
Dielectric constant at zero field was measured at a frequency of 10~kHz by using an LCR meter (E4980A, Agilent) in a PPMS.
Dielectric constant along the field direction ($E \parallel B$) up to 48~T was measured at a frequency of 50~kHz by using a capacitance bridge (1615-A, General Radio) in a ND pulsed magnet ($\sim$36~ms duration) \cite{2020_Miy}.
Silver paste was painted on the two large surfaces of a disk-shaped sintered sample to form electrodes.
Heat capacity was measured by the thermal relaxation method in the PPMS at zero field.
All the pulsed high-field experiments were performed at ISSP.

In our DFT energy mapping \cite{2021_Yam, 2021_Hei, 2022_Her}, we worked with the full potential local orbital basis set \cite{1999_Koe} and generalized gradient approximation (GGA) type exchange and correlation functional \cite{1996_Per}.
Strong electronic correlations on Cr $3d$ orbitals were treated with DFT+$U$ corrections \cite{1995_Lie}, where we varied the onsite correlation strength $U$ and fix the Hund's rule coupling to $J_{H} = 0.72$~eV \cite{1996_Miz}.
We use a $2 \times 2 \times 1$ supercell of CuGaCr$_{4}$S$_{8}$ with $Pm$ symmetry and twelve inequivalent Cr$^{3+}$ positions to extract the exchange couplings up to the fifth NN.
\section{\label{Sec3}Basic physical properties}

\subsection{\label{Sec3_1} Structural analysis}

We first show that the crystal structure of CuGaCr$_{4}$S$_{8}$ belongs to the $F{\overline 4}3m$ space group and consists of the breathing pyrochlore lattice, similar to the previously reported {\it A}-site-ordered Cr thiospinels \cite{2018_Oka, 2022_Sha}.
The synchrotron powder XRD pattern of CuGaCr$_{4}$S$_{8}$ at room temperature is depicted by red open circles in Fig.~\ref{Fig2}.
All major peaks can be indexed to the FCC symmetry characteristic of the spinel structure.
We observe small additional peaks at $2\theta = 7.98^{\circ}$ and $29.04^{\circ}$ (insets of Fig.~\ref{Fig2}), indexed as the 200 and 640 reflections, respectively.
Again, $(hk0)$ peaks with $h + k = 4n + 2$ are forbidden for $Fd{\overline 3}m$ but allowed for $F{\overline 4}3m$.
The possibility of impurity contribution to these peaks is excluded as long as we examine the known phases in the database (ICSD).
The effect of multiple diffraction \cite{1980_Tok, 1981_Ste} is minimal in our XRD measurement using the randomly oriented powder sample.
The anisotropy of the local structure around the transition metal atoms could in principle break the forbidden rule of the $Fd{\overline 3}m$ space group without the {\it A}-site ordering \cite{2004_Sub}, while the incident x-ray energy is sufficiently large compared to the resonance energies of the composition elements to suppress this factor.

\begin{table}[t]
\centering
\renewcommand{\arraystretch}{1.2}
\caption{Structural parameters of CuGaCr$_{4}$S$_{8}$ at room temperature assuming the $F{\overline 4}3m$ space group, where Cu and Ga atoms occupy the $4a$ and $4d$ sites, respectively, and the atomic position $x$ of Cr is less than 0.375. The lattice constant is $a=9.92036(8)$~\AA. Reliability factors are $R_{\rm wp}=2.826$, $R_{\rm p}=1.755$, $R_{\rm e}=1.711$, $S=1.6518$.}
\begin{tabular}{ccccccc} \hline\hline
~ & ~ & $x$ & $y$ & $z$ & ~Occup.~ & B (\AA) \\ \hline
~~~Cu~~~ & ~~$4a$~~ & ~~0~~ & ~~0~~ & ~~0~~ & 1 & ~~0.86(10)~~~ \\
~~~Ga~~~ & ~~$4d$~~ & ~~3/4~~ & ~~3/4~~ & ~~3/4~~ & 1 & ~~0.83(8)~~~ \\
~~~Cr~~~ & ~~$16e$~~ & ~~0.37042(15)~~ & ~~$x$~~ & ~~$x$~~ & 1 & ~~0.63(2)~~~ \\
~~~S1~~~ & ~~$16e$~~ & ~~0.13316(26)~~ & ~~$x$~~ & ~~$x$~~ & 1 & ~~0.79(5)~~~ \\
~~~S2~~~ & ~~$16e$~~ & ~~0.61679(21)~~ & ~~$x$~~ & ~~$x$~~ & 1 & ~~0.65(5)~~~ \\ \hline\hline
\end{tabular}
\label{model1}
\end{table}

Using the present XRD data, we performed the Rietveld analysis assuming several types of structural models.
Details of the analysis are found in Appendix~\ref{SecA}.
We confirm that the $F{\overline 4}3m$ model, assuming perfect {\it A}-site ordering and allowing the breathing distortion of the Cr pyrochlore lattice, yields better refinement than the $Fd{\overline 3}m$ model with random distribution of Cu and Ga atoms.
The fitting result and structural parameters obtained for the $F{\overline 4}3m$ space group are shown in Fig.~\ref{Fig2} and Table~\ref{model1}, respectively.
The lattice constant is $a=9.92036(8)~\AA$, in accord with Ref.~\cite{1970_Pin}.
The NN Cr--Cr bond lengths are $r=3.377(4)~\AA$ and $r'=3.638(4)~\AA$ in small and large tetrahedra, respectively.
The breathing ratio $r'/r=1.077$ is larger than those in oxides, 1.035 (LiGaCr$_{4}$O$_{8}$) and 1.049 (LiInCr$_{4}$O$_{8}$) \cite{2013_Oka}, and comparable to those in other sulfides, 1.074 (LiGaCr$_{4}$S$_{8}$), 1.089 (LiInCr$_{4}$S$_{8}$) \cite{2018_Oka}, 1.066 (CuAlCr$_{4}$S$_{8}$) \cite{2022_Sha}, and 1.084 (CuInCr$_{4}$S$_{8}$) \cite{2022_Gen}.
Although it is difficult to evaluate the ratio of site mixing between Cu and Ga, the refined value of $r'/r$ is robust to the incorporation of the site mixing in the Rietveld analysis.
We hence infer that Cu and Ga atoms are almost perfectly ordered in CuGaCr$_{4}$S$_{8}$, like in Li(Ga, In)Cr$_{4}$S$_{8}$ and Cu(Al, In)Cr$_{4}$S$_{8}$ \cite{2018_Oka, 2022_Sha}.

\begin{figure}[t]
\centering
\includegraphics[width=0.85\linewidth]{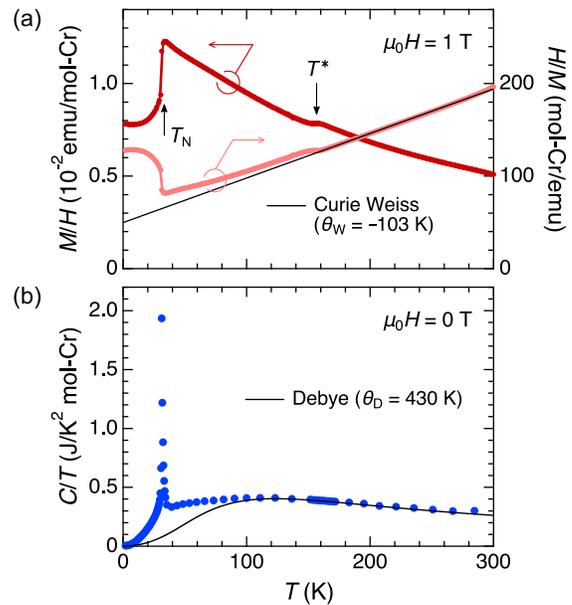}
\caption{Temperature dependence of (a) magnetic susceptibility $M/H$ at 1 T and (b) heat capacity divided by temperature $C/T$ at 0 T. The inverse magnetic susceptibility $H/M$ and the Curie-Weiss fit above 160~K (black) are displayed in the right axis of (a). The solid line in (b) denotes the estimated lattice heat capacity based on the Debye model with the Debye temperature of $\Theta_{\rm D}=430$~K.}
\label{Fig3}
\end{figure}

\subsection{\label{Sec3_2} Magnetic and structural transitions at low temperatures}

Figures~\ref{Fig3}(a) and \ref{Fig3}(b) show the magnetic susceptibility $M/H$ measured at 1 T and the heat capacity divided by temperature $C/T$ measured at 0 T as a function of temperature, respectively.
The inverse susceptibility $H/M$ exhibits linear temperature dependence between 160 and 300~K [right axis of Fig.~\ref{Fig3}(a)], following the Curie-Weiss law with the Weiss temperature of $\Theta_{\rm W}=-103$~K and the effective moment $p_{\rm eff}=4.08$~$\mu_{\rm B}$.
The large negative $\Theta_{\rm W}$ indicates dominant AFM exchange couplings, like Cu(Al, In)Cr$_{4}$S$_{8}$ \cite{2022_Sha, 2022_Gen} but unlike Li(Ga, In)Cr$_{4}$S$_{8}$ \cite{2018_Oka, 2018_Pok}.
The estimated $p_{\rm eff}$ is slightly larger than the theoretical value of 3.87~$\mu_{\rm B}$ expected for $S=3/2$ with quenched orbital moments, ensuring a nearly isotropic spin for CuGaCr$_{4}$S$_{8}$.
In order to estimate lattice contributions to $C/T$, we employ the Debye model with the Debye temperature of $\Theta_{\rm D}=430$~K as shown by a black line in Fig.~\ref{Fig3}(b).
The calculated curve fits well to the experimental data above $\sim$120~K.

\begin{figure}[t]
\centering
\includegraphics[width=0.8\linewidth]{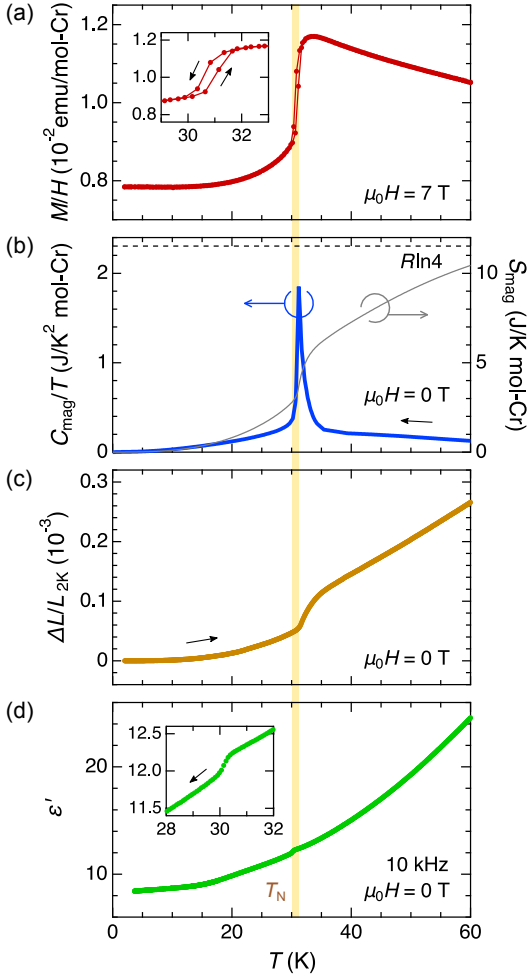}
\caption{Temperature dependence of (a) magnetic susceptibility $M/H$ at 7 T, (b) magnetic heat capacity divided by temperature $C_{\rm mag}/T$ at 0 T, (c) thermal expansion $\Delta L/L_{\rm 2K}$ at 0 T, and (d) dielectric constant $\varepsilon'$ at a frequency of 10 kHz at 0 T. The insets of (a) and (d) show enlarged views of $M/H$ and $\varepsilon'$ around $T_{\rm N}$, respectively. The magnetic entropy $S_{\rm mag}$ calculated by integrating $C_{\rm mag}/T$ with respect to temperature is displayed in the right axis of (b). Black arrows represent the direction of the temperature-sweeping process.}
\label{Fig4}
\end{figure}

In the measured temperature range, we observe several anomalies in the $M/H$-$T$ curve [Fig.~\ref{Fig3}(a)].
$M/H$ exhibits a steplike anomaly at $\sim$160~K, below which $H/M$ deviates from the linear temperature dependence.
This could be attributed to the onset of short-range AFM correlations in the main phase or to an AFM transition in an unidentified impurity phase, which we will leave open for future interpretation.
The inverse susceptibility $H/M$ further exhibits a concave behavior below $\sim$120~K, where $C/T$ deviates from the estimated lattice heat capacity [Fig.~\ref{Fig3}(b)], indicating the development of a magnetic short-range order.
The similar feature was also observed for CuInCr$_{4}$S$_{8}$ \cite{2018_Oka, 2022_Gen}.
On further cooling, an abrupt $M/H$ drop and a sharp $C/T$ peak are observed at $T_{\rm N}=31$~K, indicating the onset of a magnetic LRO.
The transition temperature is consistent with the previously reported values \cite{1970_Pin, 2005_Kes, 2018_Ami}.

\begin{figure}[t]
\centering
\includegraphics[width=0.95\linewidth]{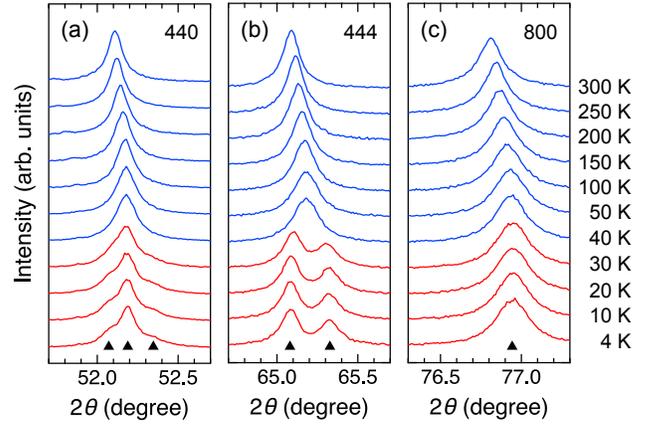}
\caption{Waterfall plots of the temperature evolution of the powder XRD pattern focusing on (a) 440, (b) 444, and (c) 800 reflections indexed for the cubic $F{\overline 4}3m$ space group. The data were obtained using a laboratory x-ray diffractometer with monochromatized Cu-$K\alpha$1 radiation. Triangles denote the peak positions obtained by the (multi-) Lorentzian fit to each peak profile.}
\label{Fig5}
\end{figure}

Figure~\ref{Fig4} summarizes the temperature dependence of various physical quantities in the vicinity of $T_{\rm N}$.
As shown in the inset of Fig.~\ref{Fig4}(a), $M/H$ shows a clear hysteresis, indicating that the magnetic transition is first order.
Figure~\ref{Fig4}(b) shows the magnetic heat capacity divided by temperature $C_{\rm mag}/T$, which is obtained by subtracting the estimated lattice contributions from the experimental $C/T$ as shown in Fig.~\ref{Fig3}(b).
By integrating $C_{\rm mag}/T$ with respect to temperature, the magnetic entropy $S_{\rm mag}$ is found to reach $\sim$5~J/(K~mol-Cr) just above $T_{\rm N}$ \cite{comment}.
We obtain $S_{\rm mag}$ at 120~K to be $\sim$13~J/(K~mol-Cr) (not shown), which roughly agrees with the theoretical value $R{\rm ln}4$ = 11.5~J/(K~mol-Cr) for the $S = 3/2$ spin system.
Notably, the thermal expansion $\Delta L/L_{\rm 2K}$ rapidly decreases below 35~K with decreasing temperature, suggesting a significant volume contraction across $T_{\rm N}$.
This behavior is similar to Li(Ga,In)Cr$_{4}$O$_{8}$ \cite{2020_Kan}, where a crystal symmetry lowering was observed \cite{2015_Nil, 2016_Sah}.
As shown below, we confirm the crystal symmetry lowering for CuGaCr$_{4}$S$_{8}$ from the powder XRD measurement at low temperatures.
Moreover, the dielectric constant $\varepsilon'$ exhibits a steplike anomaly at $T_{\rm N}$ [inset of Fig.~\ref{Fig4}(d)] as observed in Li(Ga,In)Cr$_{4}$O$_{8}$ \cite{2016_Lee, 2016_Sah}.
A previous powder neutron diffraction study \cite{1976_Wil} proposes that the magnetic structure below $T_{\rm N}$ is an incommensurate spiral state.
Thus, the observed dielectric anomaly would be of magnetic origin; in other words, CuGaCr$_{4}$S$_{8}$ would be a type-II multiferroic \cite{2009_Kho}.

\begin{figure}[t]
\centering
\includegraphics[width=0.8\linewidth]{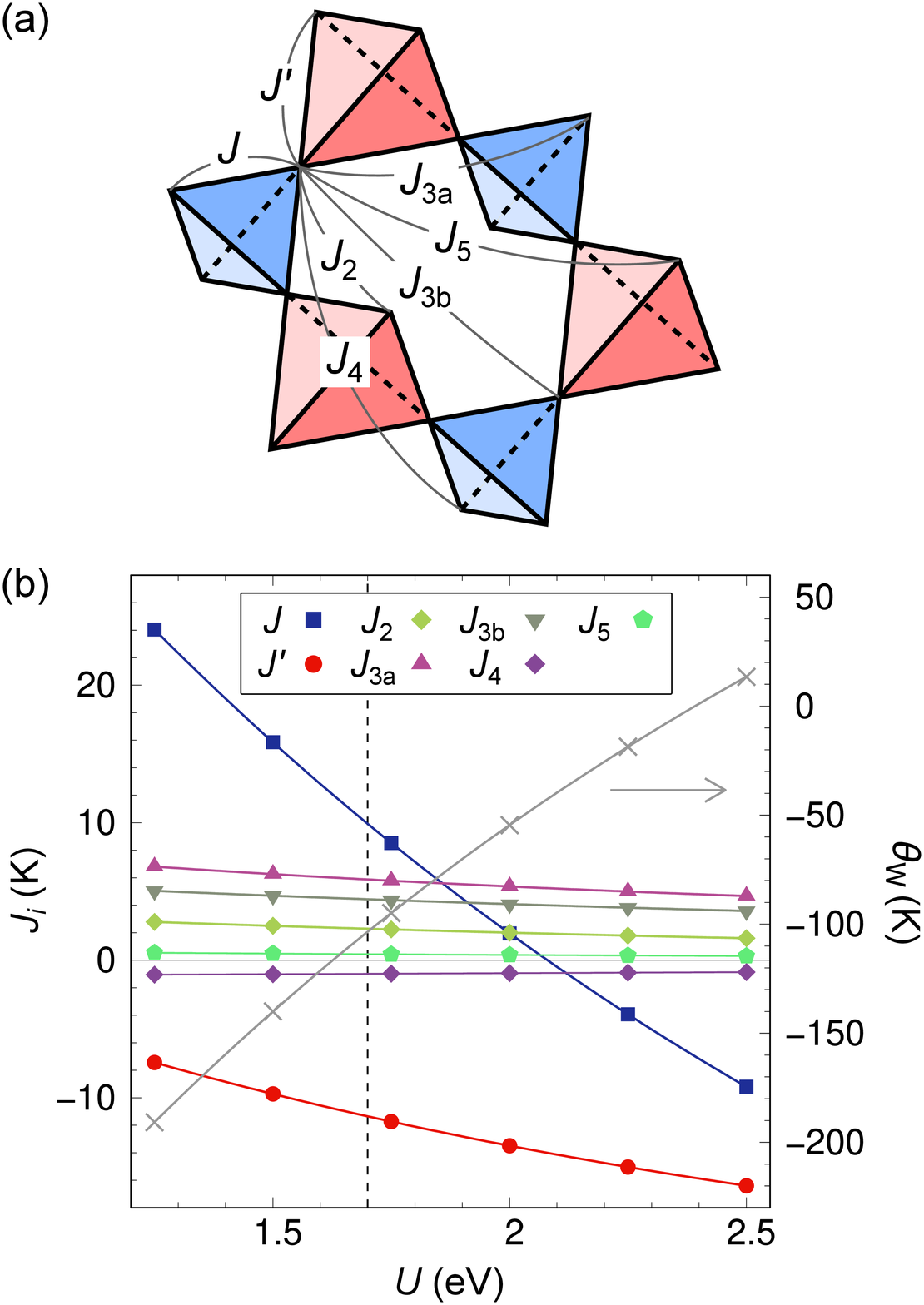}
\caption{(a) Definition of the exchange couplings up to the fifth-NN path in the breathing pyrochlore lattice. (b) Exchange parameters of CuGaCr$_{4}$S$_{8}$ at room temperature obtained by the DFT energy mapping as function of the onsite interaction strength $U$. The corresponding Weiss temperature is denoted by crosses in the right axis. The vertical line indicates the $U$ value where the exchange couplings match the experimental Weiss temperature $\Theta_{\rm W} = -103$~K.}
\label{Fig6}
\end{figure}

To reveal the structural change across the magnetic transition, we investigate the powder XRD patterns at low temperatures.
Figure~\ref{Fig5} shows the temperature evolution of peak profile of some Bragg reflections.
Peak splitting is clearly observed for many reflections below $T_{\rm N}$, signaling that the magnetic transition accompanies a structural transition.
We note that the peak splitting was not observed in the previous powder neutron diffraction study \cite{1976_Wil} because the expected splitting is much smaller than the experimental peak width, $\sim$0.2$^{\circ}$ and $\sim$2$^{\circ}$, respectively, which is not sufficient to resolve the symmetry lowering below $T_{\rm N}$.
Remarkably, $hhh$ reflections split into two peaks whereas no splitting or broadening is observed for $h00$ reflections [Figs.~\ref{Fig5}(b) and \ref{Fig5}(c)].
These observations can be accounted for by rhombohedral distortion, as opposed to tetragonal or orthorhombic distortion in other Cr spinels \cite{2006_Ued, 2007_Lee, 2009_Yok, 2009_Kan, 2015_Nil, 2016_Sah}.
However, the proposed magnetic modulation vector ${\mathbf Q} = (0.18, 0, 0.80)$ \cite{1976_Wil} may not directly cause rhombohedral distortion, but is compatible with monoclinic distortion.
Indeed, the peak profile of 440 reflection below $T_{\rm N}$ can be well fitted by the superposition of three Lorentzian functions rather than two [see Fig.~\ref{Fig5}(a) and Appendix~\ref{SecB}].
This suggests that the crystal symmetry below $T_{\rm N}$ is lower than rhombohedral.
More detailed crystallographic and magnetic structure analysis is necessary to settle the issue.

\begin{table}[t]
\centering
\renewcommand{\arraystretch}{1.2}
\caption{Exchange couplings up to third-NN in four kinds of {\it A}-site-ordered Cr thiospinels estimated by the DFT energy mapping.}
\begin{tabular}{ccccccc} \hline\hline
& $J/k_{\rm B} $ & $J'/k_{\rm B} $ & $J_{2}/k_{\rm B} $ & $J_{3a}/k_{\rm B} $ & $J_{3b}/k_{\rm B} $ & Ref. \\
\hline
LiGaCr$_{4}$S$_{8}$~ & ~$-7.7$~K~ & ~$-12.2$~K~ & ~1.2~K~ & ~6.1~K~ & ~3.0~K~ & ~\cite{2019_Gho} \\
LiInCr$_{4}$S$_{8}$~ & ~$-0.3$~K~ & ~$-28.0$~K~ & ~0.7~K~ & ~5.3~K~ & ~2.4~K~ & ~\cite{2019_Gho} \\
CuGaCr$_{4}$S$_{8}$~ & ~9.8~K~ & ~$-11.4$~K~ & ~2.3~K~ & ~5.9~K~ & ~4.4~K~ & ~This work \\
CuInCr$_{4}$S$_{8}$~ & ~14.7~K~ & ~$-26.0$~K~ & ~1.1~K~ & ~6.4~K~ & ~4.5~K~ & ~\cite{2019_Gho} \\ \hline\hline
\end{tabular}
\label{tab:J}
\end{table}

\subsection{\label{Sec3_3} DFT energy mapping}

Based on the {\it A}-site-ordered crystal structure at room temperature shown in Table~\ref{model1}, we performed the DFT calculations to estimate the exchange couplings of CuGaCr$_{4}$S$_{8}$.
More details of the calculations are found in Appendix.~\ref{SecC}.
Figure~\ref{Fig6}(b) shows the DFT energy mapping up to the fifth-NN exchange couplings, which are defined in the Heisenberg Hamiltonian of the form ${{\mathcal{H}}}=\sum_{i<j}J_{ij}{\mathbf S}_{i} \cdot {\mathbf S}_{j}$ [see Fig.~\ref{Fig6}(a)].
Exchange interactions monotonically evolve with onsite Coulomb interaction strength $U$.
The vertical dashed line indicates the $U$ value for which the exchange couplings match the experimental Weiss temperature of $\Theta_{\rm W} = -103$~K (see also Appendix~\ref{SecA}).
The obtained parameter set is $J/k_{\rm B} = 9.8(7)$~K, $J'/k_{\rm B} = -11.4(6)$~K, $J_{2}/k_{\rm B}  = 2.3(5)$~K, $J_{3a}/k_{\rm B}  = 5.9(3)$~K, $J_{3b}/k_{\rm B}  = 4.4(3)$~K, $J_{4}/k_{\rm B} = -1.0(3)$~K, and $J_{5}/k_{\rm B}  = 0.4(3)$~K, where $k_{\mathrm{B}}$ is the Boltzmann's constant, representing that CuGaCr$_{4}$S$_{8}$ is characterized by strong AFM $J$ and FM $J'$.
Since the strengths of the exchange couplings should be modified below $T_{\rm N}$ due to the lower-symmetry crystal structure, the ground state cannot be simply represented by these parameters.
Nevertheless, our magnetization measurements suggest that the low-temperature magnetism of CuGaCr$_{4}$S$_{8}$ can be understood based on the AFM-$J$-FM-$J'$ picture, as discussed in Sec.~\ref{Sec4_1}.

Table~\ref{tab:J} compares the exchange parameters among four kinds of {\it A}-site-ordered Cr thiospinels estimated in the previous works \cite{2019_Gho} and this work.
One can find that $J$ and $J'$ are strongly dependent on the  types of nonmagnetic cations.
The occupation of Li atoms at the $4a$ site leads to FM $J$, whereas that of Cu atoms leads to AFM $J$.
$J'$ is always FM, and its strength is enhanced when In atoms occupy the $4d$ site.
These tendencies are reasonable because monovalent {\it A}$^{+}$ (trivalent {\it A'}$^{3+}$) cations are surrounded by S1 (S2) atoms connecting the short (long) NN Cr--Cr bonds [Fig.~\ref{Fig1}(b)].
The Hamiltonian of CuGaCr$_{4}$S$_{8}$ is qualitatively similar to that of CuInCr$_{4}$S$_{8}$ except that $|J'|$ is much smaller.
In other words, the effects of further-neighbor interactions, especially $J_{3a}$ and $J_{3b}$, are more important for CuGaCr$_{4}$S$_{8}$ than for CuInCr$_{4}$S$_{8}$.
This may be responsible for the difference in the ground state at zero field; a commensurate ${\mathbf Q} = (1, 0, 0)$ state with an $S=6$ spin cluster in the large tetrahedron is realized for CuInCr$_{4}$S$_{8}$ \cite{2021_Gao}, whereas an incommensurate spiral state in which four spins in the large tetrahedra are not parallel with each other for CuGaCr$_{4}$S$_{8}$ \cite{1976_Wil}.

\section{\label{Sec4} Magnetic-field induced phase transitions}

\subsection{\label{Sec4_1} Magnetization curves}

\begin{figure}[t]
\centering
\includegraphics[width=\linewidth]{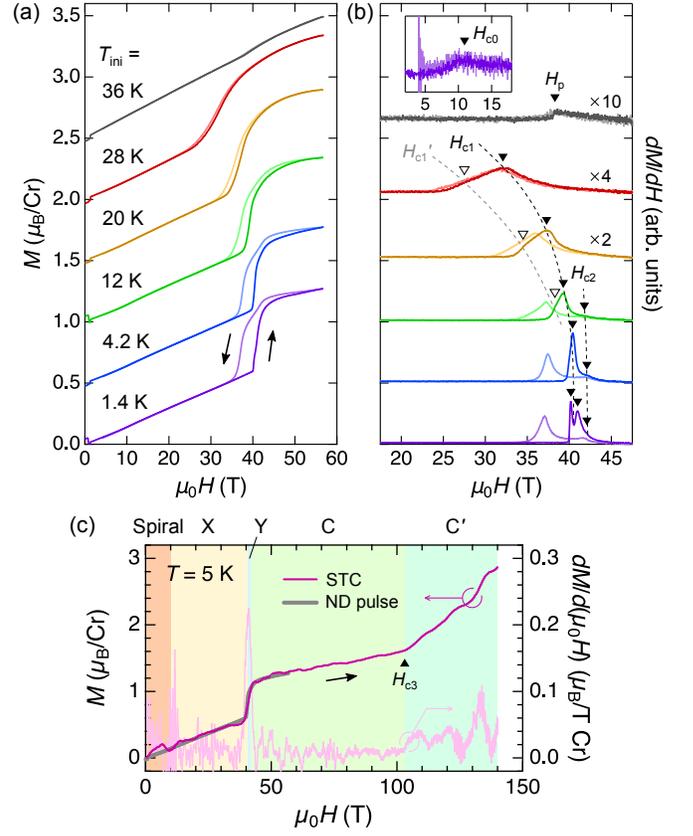}
\caption{(a)(b) Magnetic-field dependence of (a) magnetization $M$ and (b) its field derivative $dM/dH$ measured at various initial temperatures $T_{\rm ini}$ in a non-destructive (ND) pulsed magnet. Thick (thin) lines correspond to the data in field-increasing (decreasing) processes. The curves except for $T_{\rm ini}=1.4$~K are shifted upward for clarity. The inset of (b) is an enlarged view of $dM/dH$ around the lowest-field phase transition for $T_{\rm ini}=1.4$~K. (c) Magnetic-field dependence of $M$ (left) and $dM/dH$ (right) in the field-increasing process measured at $T_{\rm ini} \sim 5$~K in a single-turn-coil (STC) system. The absolute value of $M$ is calibrated by fitting with the $M$--$H$ curve for $T_{\rm ini} = 4.2$~K obtained in a ND pulsed magnet (gray).}
\label{Fig7}
\end{figure}

We here move on to the field-induced properties of CuGaCr$_{4}$S$_{8}$ revealed by pulsed high-field experiments.
Figure~\ref{Fig7}(a) shows magnetization curves measured at various initial temperatures $T_{\rm ini}$ using the ND pulsed magnet.
Although they appear simple at a glance, anomalies in the field derivatives $dM/dH$ show a complicated temperature dependence, as shown in Fig.~\ref{Fig7}(b).
For $T_{\rm ini}=4.2$~K, a drastic magnetization jump is observed at around 40~T accompanied by a large hysteresis, where $dM/dH$ exhibits a sharp peak at $\mu_{0}H_{\rm c1}=40.4$~T and a shoulder-like anomaly at $\mu_{0}H_{\rm c2}=42.1$~T in the field-increasing process.
The shoulder-like anomaly does not change its position while the peak moves to a lower field at 37.4~T in the field-decreasing process.
After the metamagnetic transition, $M$ reaches $\sim$1.2~$\mu_{\rm B}$/Cr, which is much smaller than the expected saturation value of $\sim$3~$\mu_{\rm B}$/Cr.
As $T_{\rm ini}$ increases, $H_{\rm c1}$ shifts to a lower field and the hysteresis becomes smaller.
Notably, a broad shoulder-like structure appears on the low-field side of the $dM/dH$ peak at $H_{\rm c1}'$, as denoted by open triangles in Fig.~\ref{Fig7}(b).
As shown in Sec.~\ref{Sec4_2}, this magnetization anomaly is accompanied by a pronounced dielectric response.
Even for $T_{\rm ini}=36$~K ($>T_{\rm N}$), a weak metamagnetic transition from the paramagnetic phase is observed at $\mu_{0}H_{\rm p}=38.3$~T, indicating that the field-induced phase is robust against thermal fluctuations.

For all the measured $T_{\rm ini}$'s below $T_{\rm N}$, a subtle slope change in the magnetization curve is observed at $\mu_{0}H_{\rm c0} \approx 10$~T, which is visible as a $dM/dH$ cusp [inset of Fig.~\ref{Fig7}(b)].
This phase transition is also confirmed by the magnetization measurement up to 14~T in a static magnetic field (see Fig.~\ref{Fig11} in Appendix~\ref{SecD}).
We ascribe these anomalies to a spin-flop transition with the reorientation of magnetic domains of the helical state, as observed for CdCr$_{2}$O$_{4}$ \cite{2007_Mat_1, 2011_Bha, 2019_Ros, 2020_Ros, 2005_Chu}.
For $T_{\rm ini}=1.4$~K, $dM/dH$ exhibits two peaks around 40~T in the field-increasing process [Fig.~\ref{Fig7}(b)].
We tentatively view this behavior as a splitting of the peak at $H_{\rm c1}$ observed for $T_{\rm ini} \geq 4.2$~K, though its origin is unclear at present.

\begin{table}[t]
\centering
\renewcommand{\arraystretch}{1.2}
\caption{Critical fields of the successive phase transitions in CuGaCr$_{4}$S$_{8}$ and CuInCr$_{4}$S$_{8}$ at $\sim$5~K. $H_{\rm c0}$ indicates a spin-flop transition in a low-field region observed only for CuGaCr$_{4}$S$_{8}$. $H_{\rm c1} \sim H_{\rm c3}$ correspond to the termination fields of {\it X}, {\it Y}, and {\it C} phases, which can be assigned to canted 2:2, canted 2:1:1, and 3-up-1-down phases, respectively, based on the magnetoelastic theory \cite{2020_Gen}. $H_{\rm sat}$ indicates the expected saturation field deduced from Eq.~(\ref{eq:H_sat}) ($J'$: FM) using the exchange couplings based on the DFT calculations \cite{2019_Gho}.}
\begin{tabular}{ccccccc} \hline\hline
& $\mu_{0}H_{\rm c0}$ & $\mu_{0}H_{\rm c1}$ & $\mu_{0}H_{\rm c2}$ & $\mu_{0}H_{\rm c3}$ & $\mu_{0}H_{\rm sat}$ & Ref. \\
\hline
CuGaCr$_{4}$S$_{8}$~ & ~10.8~T~ & ~40.4~T~ & ~42.1~T~ & ~103~T~ & ~$170$~T~ & ~This work \\
CuInCr$_{4}$S$_{8}$~ & ~---~ & ~32~T~ & ~56~T~ & ~112~T~ & ~180~T~ & ~\cite{2020_Gen, 2022_Gen} \\ \hline\hline
\end{tabular}
\label{tab:MH}
\end{table}

To get a whole picture of the field-induced phase transitions, we further measure the magnetization at $T_{\rm ini} \sim 5$~K up to $\sim$140~T using the STC system.
As shown in Fig.~\ref{Fig7}(c), a plateau-like behavior is observed between $\mu_{0}H_{\rm c1}=40$~T and $\mu_{0}H_{\rm c3}=103$~T in the field-increasing process.
Note that the transition at $H_{\rm c2}$ is not resolved due to an electromagnetic noise.
Judging from the magnitude of $M$ in this field range ($\sim$1.5~$\mu_{\rm B}$/Cr), a 3-up-1-down state with a 1/2-magnetization plateau is expected to appear as in Cr spinel oxides \cite{2006_Ued, 2008_Koj, 2011_Miy_JPSJ, 2014_Miy, 2019_Gen, 2007_Mat_2, 2010_Mat} and CuInCr$_{4}$S$_{8}$ \cite{2020_Gen}.
Above $H_{\rm c3}$, $M$ rapidly increases up to the applied maximum field of 140~T, where $M$ reaches $\sim$2.8~$\mu_{\rm B}$/Cr.
This indicates that the saturation field $H_{\rm sat}$ is a bit higher than 140~T.

Theoretically, AFM-$J$-AFM-$J'$ or AFM-$J$-FM-$J'$ breathing pyrochlore magnets with the spin-lattice coupling can host a 1/2-magnetization plateau \cite{2021_Aoy, 2020_Gen}.
If we consider the mean-field approximation and neglect the spin-lattice coupling, $\Theta_{\rm W}$ and $H_{\rm sat}$ are related to the exchange couplings as
\vspace{-0.1cm}
\begin{equation}
\begin{split}
\label{eq:Weiss}
-\frac{S(S+1)}{k_{\rm B}}\Theta_{\rm W}=J+J'+{\overline J}_{\rm AFM}+{\overline J}_{\rm FM}
\end{split}
\end{equation}
and
\begin{equation}
\label{eq:H_sat}
\frac{g\mu_{\rm B}}{4S}\mu_{0}H_{\rm sat} = \left\{
\begin{array}{ll}
J+J'+{\overline J}_{\rm AFM} & (J': {\rm AFM})\\
J+{\overline J}_{\rm AFM} & (J': {\rm FM}),
\end{array}
\right.
\end{equation}
where ${\overline J}_{\rm AFM}$ and ${\overline J}_{\rm FM}$ are the summation of AFM and FM further-neighbor exchange couplings, respectively, and $g \approx 2.1$ is the Land\'{e} $g$ factor estimated from the Curie-Weiss fit [Fig.~\ref{Fig3}(a)].
For CuGaCr$_{4}$S$_{8}$, if we assume AFM-$J$-AFM-$J'$ and ${\overline J}_{\rm FM} \approx 0$, the saturation field is estimated to $\mu_{0}H_{\rm sat} \approx 117$~T.
This value is significantly underestimated compared to the experimentally expected value ($> 140$~T), ensuring that the AFM-$J$-FM-$J'$ picture is appropriate, as suggested by the DFT calculations (Sec.~\ref{Sec3_1}).
If we substitute the exchange parameters shown in Table~\ref{tab:J} into Eq.~(\ref{eq:H_sat}), where ${\overline J}_{\rm AFM} = 4J_{2}+2J_{3a}+2J_{3b}+2J_{5}$ and ${\overline J}_{\rm FM}=2J_{4}$, the saturation field is estimated to $\mu_{0}H_{\rm sat} \approx 170$~T.
This seems to be in good agreement with the experimental magnetization curve at 5~K [Fig.~\ref{Fig7}(c)], given that $H_{\rm sat}$ would be suppressed by the spin-lattice coupling and thermal fluctuations.

The overall magnetization curve of CuGaCr$_{4}$S$_{8}$ at 4.2 or 5~K is similar to that of CuInCr$_{4}$S$_{8}$ \cite{2020_Gen, 2022_Gen}.
For both the compounds, an intermediate-field phase appears just below the 1/2-magnetization plateau.
The magnetic structure in the intermediate-field phase is expected to be a canted 2:1:1 state according to the magnetoelastic theory assuming the effective $S=6$ FCC-lattice model \cite{2020_Gen}, which predicts successive phase transitions from a canted 2:2 to canted 2:1:1, 3-up-1-down, canted 3:1, and a fully polarized phase.
Note that a canted 2:1:1 phase is also observed for ZnCr$_{2}$O$_{4}$ \cite{2011_Miy_JPSJ} and MgCr$_{2}$O$_{4}$ \cite{2014_Miy} in a narrow field range.
An incommensurate spiral component would coexist in the canted spin states in CuGaCr$_{4}$S$_{8}$ due to the presence of sizable further-neighbor exchange couplings and/or the DM interaction, as proposed for CdCr$_{2}$O$_{4}$ \cite {2007_Mat_1, 2020_Ros}.
In the following, we call the field-induced phases of CuGaCr$_{4}$S$_{8}$ at 5~K {\it X}, {\it Y}, {\it C}, and {\it C'} phases in the ascending order of the field [Fig.~\ref{Fig7}(c)].
Table~\ref{tab:MH} summarizes the critical field of each phase transition in the field-increasing process for CuGaCr$_{4}$S$_{8}$ and CuInCr$_{4}$S$_{8}$ \cite{2020_Gen, 2022_Gen}.
Theoretically, as the SLC gets stronger, the 1/2-magnetization plateau expands while the canted 2:1:1 phase gets narrower \cite{2020_Gen}.
Considering $H_{\rm c3}/H_{\rm c2} \sim 2.5$ and 2.0 for CuGaCr$_{4}$S$_{8}$ and CuInCr$_{4}$S$_{8}$, respectively, it can be said that the 3-up-1-down state is more stable in CuGaCr$_{4}$S$_{8}$ than in CuInCr$_{4}$S$_{8}$.
Besides, the field range between $H_{\rm c1}$ and $H_{\rm c2}$ ({\it Y} phase) is much narrower in CuGaCr$_{4}$S$_{8}$ than in CuInCr$_{4}$S$_{8}$.
These trends suggest that the SLC is stronger in CuGaCr$_{4}$S$_{8}$ than in CuInCr$_{4}$S$_{8}$.
The strong SLC in CuGaCr$_{4}$S$_{8}$ is consistent with the observation of a first-order phase transition at $T_{\rm N}$.

\begin{figure}[t]
\centering
\includegraphics[width=\linewidth]{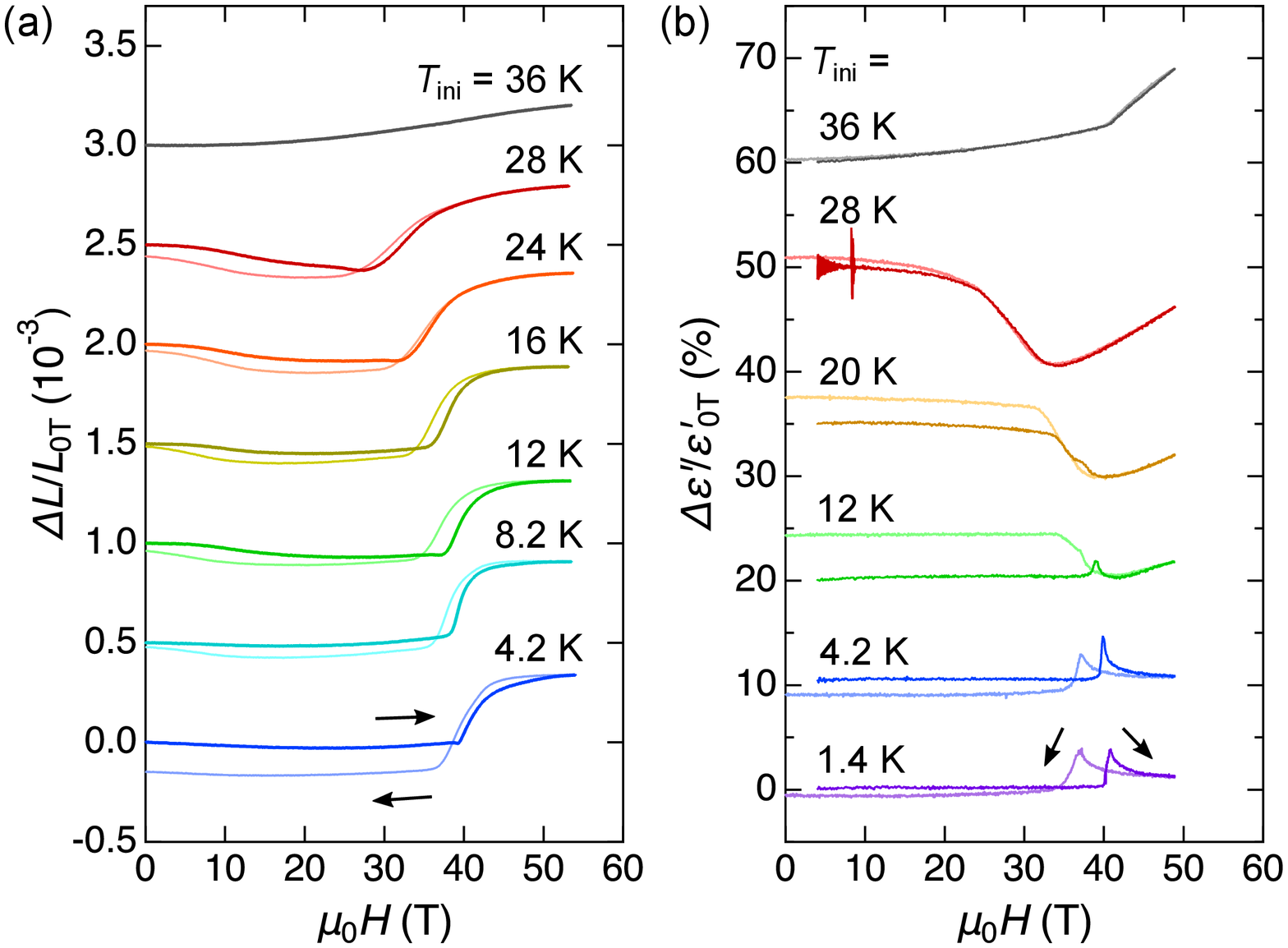}
\caption{Magnetic-field dependence of relative changes in (a) the sample length along the field direction $\Delta L/L_{\rm 0T}$ and (b) dielectric constant $\Delta \varepsilon'/\varepsilon'$ measured at various initial temperatures $T_{\rm ini}$. Data presentation is the same as Fig.~\ref{Fig7}.}
\label{Fig8}
\end{figure}

\subsection{\label{Sec4_2} Magnetostrictive and magnetodielectric effects}

Figure~\ref{Fig8}(a) shows the field dependence of the longitudinal magnetostriction and the longitudinal dielectric constant, respectively, measured on a sintered sample at various $T_{\rm ini}$'s using the ND pulsed magnet. 
Here, $\Delta L/L_{\rm 0T}$ and $\Delta \varepsilon'/\varepsilon'_{\rm 0T}$ represent the relative changes normalized by the zero-field values at each $T_{\rm ini}$.
For $T_{\rm ini}=36$~K, $\Delta L/L_{\rm 0T}$ shows a parabolic field dependence at low fields as expected in the paramagnetic state.
For $T_{\rm ini}$ below $T_{\rm N}$, on the other hand, $\Delta L/L_{\rm 0T}$ remains almost constant or exhibits negative magnetostriction in the spiral phase and {\it X} phase.
This tendency is in contrast to Li(Ga, In)Cr$_{4}$S$_{8}$ \cite{2020_Kan} and CuInCr$_{4}$S$_{8}$ \cite{2022_Gen}, where $\Delta L/L_{\rm 0T}$ gradually increases with respect to a magnetic field.
A slight slope change at around 10~T observed for $12 \leq T_{\rm ini} \leq 28$~K would reflect the spin-flop transition at $H_{\rm c0}$.
On entering the 1/2-magnetization plateau above $H_{\rm c1}$, a drastic lattice expansion is observed in analogy with other Cr spinels \cite{2022_Gen, 2007_Tan, 2019_Ros}, suggesting that the 3-up-1-down collinear state is stabilized by the exchange striction.

Interestingly, diverse dielectric responses are observed as shown in Fig.~\ref{Fig8}(b).
For $T_{\rm ini}=1.4$ and 4.2~K, $\Delta \varepsilon'/\varepsilon'_{\rm 0T}$ shows a sharp peak with a magnitude of $\sim$4\% between {\it X} and {\it Y} phases at $H_{\rm c1}$.
As $T_{\rm ini}$ increases toward $T_{\rm N}$, the dielectric anomaly around $H_{\rm c1}$ transforms into a broad valley shape with a reduction of as large as $\sim$8\%.
For $T_{\rm ini}=12$~K, a valley structure is seen only in the field-decreasing process, presumably suggesting that the actual sample temperature would be higher than $T_{\rm ini}$ in the field-decreasing process due to the magnetocaloric effect \cite{2022_Gen, 2022_Kim}.
Of particular note is the 20-K data, in which a tiny $\Delta \varepsilon'/\varepsilon'$ peak coexist with the valley structure.
This supports the occurrence of another phase transition other than that from {\it X} to {\it Y} phase, as suggested by the magnetization data [Fig.~\ref{Fig7}(b)].
We hereafter call the additional higher-temperature phase ``{\it Z} phase".

The magnetoelectric effect of the {\it A}-site-ordered spinel has long been a subject of interest in view of the inversion symmetry breaking of the crystal structure \cite{2014_Ter, 2021_Sun}.
However, the reported dielectric anomalies associated with magnetic transitions are weak for Li(Ga, In)Cr$_{4}$O$_{8}$ \cite{2016_Lee, 2016_Sah}, where a collinear 2-up-2-down magnetic LRO is induced by the SLC at low temperatures.
We stress that the observed dielectric anomaly in CuGaCr$_{4}$S$_{8}$ is much larger than in the oxide cases thanks to the emergence of incommensurate magnetic LROs \cite{1976_Wil}.
A single-crystal study on CuGaCr$_{4}$S$_{8}$ would be a promising route to seek for significant electric polarization changes.

\begin{figure}[t]
\centering
\includegraphics[width=0.75\linewidth]{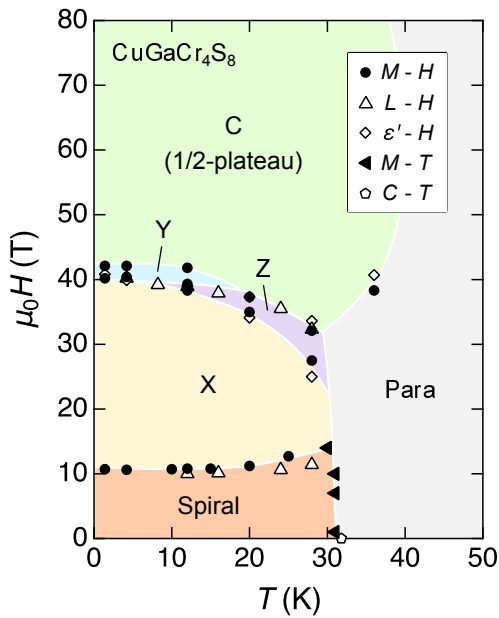}
\caption{$H$-$T$ phase diagram of CuGaCr$_{4}$S$_{8}$ based on physical property measurements performed in the present work. Note that another high-field phase ({\it C'} phase) appears above $\mu_{0}H_{\rm c3} = 103$~T at $\sim$5~K. According to the magnetoelastic theory \cite{2020_Gen}, the magnetic structure of {\it C} phase is expected to be a 3-up-1-down state.}
\label{Fig9}
\end{figure}

\subsection{\label{Sec4_3} {\textit H}-{\textit T} phase diagram}

Based on a series of physical property measurements, we construct an $H$-$T$ phase diagram of CuGaCr$_{4}$S$_{8}$, as shown in Fig.~\ref{Fig9}.
As discussed in Sec.~\ref{Sec4_1}, the magnetic structures of {\it X}, {\it Y}, {\it C} phases are possibly canted 2:2, canted 2:1:1, and 3-up-1-down states, respectively, based on the magnetoelastic theory \cite{2020_Gen}, though in reality an incommensurate spiral component would coexist with the commensurate collinear component.
We also observe a weak spin-flop transition at around 10~T, so that the {\it X} phase may be just a flopped spiral state with ${\mathbf Q} = (0.18, 0, 0.80)$ \cite{1976_Wil}.
An additional magnetization measurement using the STC system reveals that the {\it C} phase terminates at $\mu_{0}H_{\rm c3} = 103$~T, followed by the {\it C'} phase with a possible canted 3:1 state up to at least 140~T at $\sim$5~K.

The identified magnetic phase diagram is basically shared with those of Cr spinel oxides \cite{2006_Ued, 2019_Ros, 2008_Koj, 2011_Miy_JPSJ, 2014_Miy, 2022_Kim} and CuInCr$_{4}$S$_{8}$ \cite{2020_Gen, 2022_Gen} where a robust 1/2-magnetization plateau ({\it C} phase) intervenes between spin-canted phases on lower- and higher-field sides.
A characteristic feature of CuGaCr$_{4}$S$_{8}$ is the appearance of a field-induced high-temperature phase ({\it Z} phase) immediately below the {\it C} phase, in common with another AFM-$J$-FM-$J'$ breathing pyrochlore compound CuInCr$_{4}$S$_{8}$, which hosts an {\it A} phase in a closed $H$-$T$ regime around 25--40 T and 10--35 K \cite{2022_Gen}.
In the case of CuInCr$_{4}$S$_{8}$, the appearance of the {\it A} phase is accompanied by negative magnetostriction and the enhancement of magnetocapacitance at a lower phase boundary \cite{2022_Gen}.
These features are not the case for the {\it Z} phase in CuGaCr$_{4}$S$_{8}$.
Thus, we infer that the {\it Z} phase in CuGaCr$_{4}$S$_{8}$ is different from the {\it A} phase in CuInCr$_{4}$S$_{8}$.
The identification of these field-induced high-temperature phases in AFM-$J$-FM-$J'$ breathing pyrochlore systems would be an intriguing issue left for future works.

\section{\label{Sec5}Summary}

We synthesized CuGaCr$_{4}$S$_{8}$ polycrystalline samples and demonstrated the formation of a breathing pyrochlore Cr network by the synchrotron XRD measurement and the Rietveld analysis.
The DFT energy mapping suggests that the spin Hamiltonian is characterized by AFM $J$, FM $J'$, and relatively strong further-neighbor exchange couplings, so that the system harbors both geometrical frustration and bond frustration.
Remarkably, the comparison of exchange parameters among four {\it AA'}Cr$_{4}$S$_{8}$ compounds (Table~\ref{tab:J}) provides insights into the relationship between the chemical composition and the spin Hamiltonian in breathing pyrochlore chromium thiospinels.
We also unveil diverse magnetic phases associated with prominent magnetostrictive and magnetodielectric effects at low temperatures and in high magnetic fields.
These observations in CuGaCr$_{4}$S$_{8}$ could be attributed to the magnetic frustration and strong SLC, highlighting the complex interplay of these factors.

\section*{Acknowledgements}
We appreciate Y. Okamoto, M. Mori, and S. Kitou for helpful discussions. 
This work was financially supported by the Japan Society for the Promotion of Science (JSPS) KAKENHI Grants-In-Aid for Scientific Research (No. 20J10988).
M.G. was a postdoctoral research fellow of the JSPS.

\appendix

\section{\label{SecA} Rietveld analysis assuming various structural models}

\begin{figure}[t]
\centering
\includegraphics[width=\linewidth]{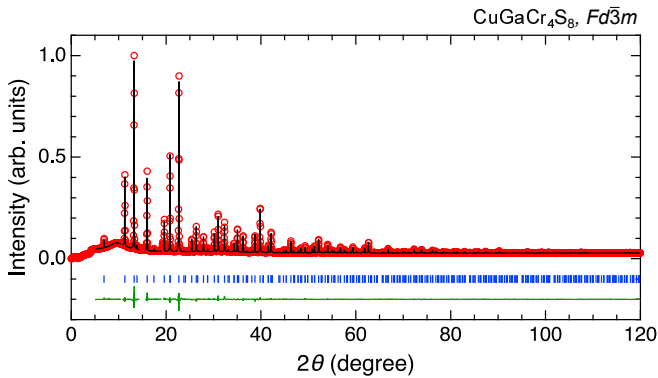}
\caption{Rietveld refinement on the synchrotron XRD pattern based on the $Fd{\overline 3}m$ space group.}
\label{Fig10}
\end{figure}

Here, we show the results of the Rietveld analysis assuming several types of structural models.
Basically, the normal spinel structure with the $Fd{\overline 3}m$ space group, in which Cu and Ga atoms are randomly distributed at the $8a$ site with an occupancy of 0.5, yields satisfactory refinement with relatively small reliability factors, as shown in Fig.~\ref{Fig10} and Table~\ref{model2}.
However, upon relaxing the constraint on the atomic position of Cr, $x({\rm Cr})$, while maintaining the random distribution of Cu and Ga and the atomic position of S, $x({\rm Cr})$ tends to deviate from 0.375, suggesting that the $F{\overline 4}3m$ space group is more likely.
Such symmetry lowering should originate from the crystallographic ordering of Cu$^{+}$ and Ga$^{3+}$ with different ionic radii (0.60~$\AA$ and 0.47~$\AA$, respectively, in the four-fold coordination).
Accordingly, we consider two types of perfectly {\it A}-site-ordered structures with $x({\rm Cr})<0.375$: Cu (Ga) occupies the $4a$ ($4d$) site for type~\#1, whereas Cu (Ga) occupies the $4d$ ($4a$) site for type~\#2.
Note that there can be two inequivalent atomic sites of S for the $F{\overline 4}3m$ space group \cite{2013_Oka, 2018_Oka, 2022_Sha}.
The refinement results for the type~\#1 and \#2 structures are shown in Tables~\ref{model1} and \ref{model3}, respectively.
In the both refinements, the reliability factors are smaller compared to the case of the $Fd{\overline 3}m$ model.
Although the resultant structural parameters are almost the same within errors between type~\#1 and \#2, the thermal parameter of Ga is unusually large for type~\#2, suggesting that type~\#1 is more reasonable.
This is further supported by the calculation of DFT total energies; type~\#1 is 293~meV per formula unit lower in energy than type~\#2.

\begin{table}[b]
\centering
\renewcommand{\arraystretch}{1.2}
\caption{Structural parameters of CuGaCr$_{4}$S$_{8}$ at room temperature assuming the $Fd{\overline 3}m$ space group, where Cu and Ga atoms are randomly distributed at the $8a$ site with an occupancy of 0.5. The lattice constant is $a=9.91910(9)$~\AA. Reliability factors are $R_{\rm wp}=2.967$, $R_{\rm p}=1.897$, $R_{\rm e}=1.706$, $S=1.7390$.}
\begin{tabular}{ccccccc} \hline\hline
~ & ~ & $x$ & $y$ & $z$ & ~Occup.~ & B (\AA) \\ \hline
~~~Cu~~~ & ~~$8a$~~ & ~~0~~ & ~~0~~ & ~~0~~ & 0.5 & ~~0.80(2)~~~ \\
~~~Ga~~~ & ~~$8a$~~ & ~~0~~ & ~~0~~ & ~~0~~ & 0.5 & ~~0.80(2)~~~ \\
~~~Cr~~~ & ~~$16d$~~ & ~~0.375~~ & ~~$x$~~ & ~~$x$~~ & 1 & ~~0.85(2)~~~ \\
~~~S~~~ & ~~$32e$~~ & ~~0.38382(6)~~ & ~~$x$~~ & ~~$x$~~ & 1 & ~~0.61(2)~~~ \\ \hline\hline
\end{tabular}
\label{model2}
\end{table}

\begin{table}[t]
\centering
\renewcommand{\arraystretch}{1.2}
\caption{Structural parameters of CuGaCr$_{4}$S$_{8}$ at room temperature assuming the $F{\overline 4}3m$ space group, where Cu and Ga atoms occupy the $4d$ and $4a$ sites, respectively, and the atomic position $x$ of Cr is less than 0.375. The lattice constant is $a=9.92035(8)$~\AA. Reliability factors are $R_{\rm wp}=2.838$, $R_{\rm p}=1.773$, $R_{\rm e}=1.711$, $S=1.6588$.}
\begin{tabular}{ccccccc} \hline\hline
~ & ~ & $x$ & $y$ & $z$ & ~Occup.~ & B (\AA) \\ \hline
~~~Cu~~~ & ~~$4d$~~ & ~~3/4~~ & ~~3/4~~ & ~~3/4~~ & 1 & ~~0.69(6)~~~ \\
~~~Ga~~~ & ~~$4a$~~ & ~~0~~ & ~~0~~ & ~~0~~ & 1 & ~~1.00(7)~~~ \\
~~~Cr~~~ & ~~$16e$~~ & ~~0.37041(14)~~ & ~~$x$~~ & ~~$x$~~ & 1 & ~~0.64(2)~~~ \\
~~~S1~~~ & ~~$16e$~~ & ~~0.13340(22)~~ & ~~$x$~~ & ~~$x$~~ & 1 & ~~0.79(4)~~~ \\
~~~S2~~~ & ~~$16e$~~ & ~~0.61683(17)~~ & ~~$x$~~ & ~~$x$~~ & 1 & ~~0.62(4)~~~ \\ \hline\hline
\end{tabular}
\label{model3}
\end{table}

\section{\label{SecB} Lorentzian fit on the peak profile of 440 reflection at low temperatures}

\begin{figure}[b]
\centering
\includegraphics[width=0.65\linewidth]{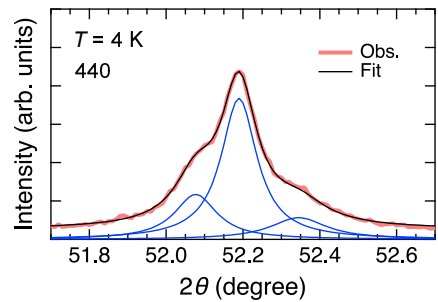}
\caption{Multi-Lorentzian fit (black) to the peak profile of 440 reflection at 4~K (pink), which is obtained by the superposition of three Lorentzian functions (blue).}
\label{Fig11}
\end{figure}

Figure~\ref{Fig11} shows an enlarged view of the powder XRD pattern of CuGaCr$_{4}$S$_{8}$ around the 440 reflection at 4~K.
This peak can be fitted by the superposition of three Lorentzian functions with peak positions of 52.08$^{\circ}$, 52.19$^{\circ}$, and 52.34$^{\circ}$.

\section{\label{SecC} Details of the DFT energy mapping}

\begin{figure}[t]
\centering
\includegraphics[width=0.8\linewidth]{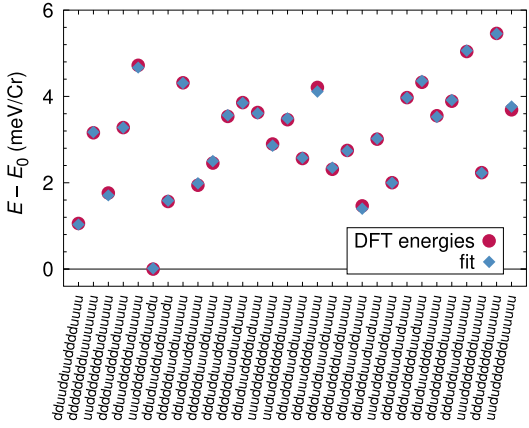}
\caption{DFT energies of CuGaCr$_{4}$S$_{8}$, calculated within GGA+$U$ at $J_{H} = 0.72$~eV, $U = 1.70$~eV and $6 \times 6 \times 6$ $k$ points in a $2 \times 2 \times 1$ supercell for 30 different spin configurations, compared with a fit to the Heisenberg model. The good fit indicates that the extracted exchange interactions represent the magnetic properties of CuGaCr$_{4}$S$_{8}$ well.}
\label{Fig12}
\end{figure}

Our DFT calculations are performed using the all electron full potential FPLO code \cite{1999_Koe}.
We create a $2 \times 2 \times 1$ supercell with $Pm$ symmetry and twelve symmetry inequivalent Cr$^{3+}$ ions.
This allows us to go slightly beyond the minimal set of exchange interactions $J$, $J'$, $J_{2}$, $J_{3a}$, and $J_{3b}$,  which is known to describe the breathing pyrochlore thiospinels well \cite{2019_Gho, 2020_Gen}; we also resolve $J_{4}$ which has been shown to noticeably affect the inelastic neutron spectrum of CuInCr$_{4}$S$_{8}$ \cite{2021_Gao} and ZnCr$_{2}$Se$_{4}$ \cite{2022_Gao}.
We calculate 30 spin configurations and fit them with the classical Heisenberg Hamiltonian energies; the good quality of the fit is shown in Fig.~\ref{Fig12}.
We select the onsite interactions strength $U$ which is suitable for the description of CuGaCr$_{4}$S$_{8}$ by calculating the Weiss temperature for the obtained exchange interactions via Eq.~(\ref{eq:Weiss}), where ${\overline J}_{\rm AFM} = 4J_{2}+2J_{3a}+2J_{3b}+2J_{5}$ and ${\overline J}_{\rm FM}=2J_{4}$.
For an interpolated $U = 1.70$~eV, the set of couplings exactly matches the experimental value $\Theta_{\rm W} = -103$~ K.
This $U$ value falls into the range of $U$ values $1.4~{\rm eV} < U < 2.0~{\rm eV}$ that describe many other chromium spinels \cite{2019_Gho}.

\section{\label{SecD} Low-field phase transition observed in a static magnetic field}

\begin{figure}[h]
\centering
\includegraphics[width=0.85\linewidth]{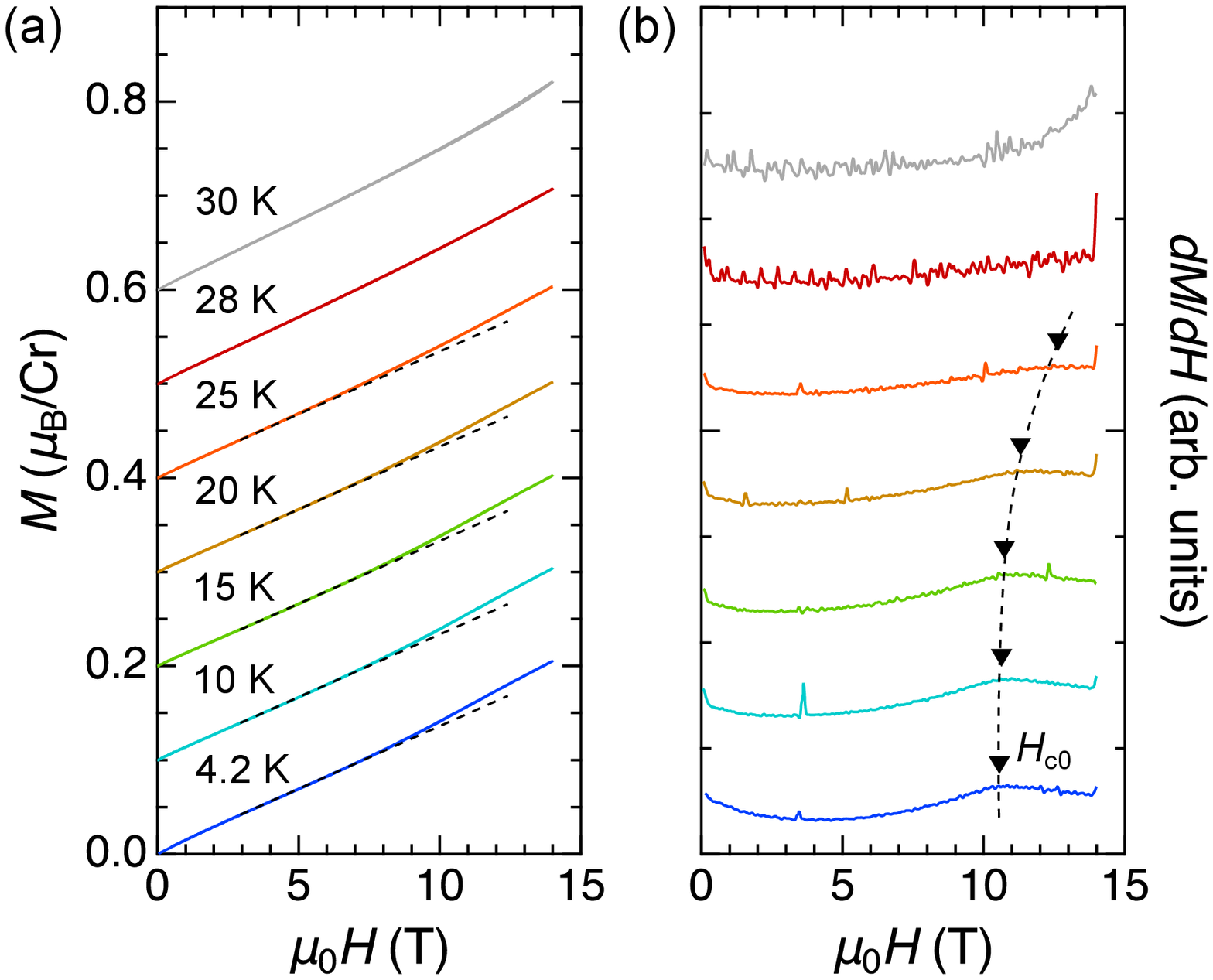}
\caption{Magnetic-field dependence of (a) magnetization $M$ and (b) its field derivative $dM/dH$ measured at various temperatures in a static magnetic field. The curves except for 4.2~K are shifted upward for clarity. Dashed lines in (a) are guides to the eye to make it easier to see the slope change in the $M$--$H$ curves.}
\label{Fig13}
\end{figure}

Figure~\ref{Fig13} shows the magnetization curves measured at various temperatures using a vibrating sample magnetometer in a PPMS.
A cusp structure is clearly observed in the $dM/dH$ data below 25~K, indicating a spin-flop transition.
The transition field $H_{\rm c0}$ gradually increases as the temperature increases.
This trend is in line with the typical $H$-$T$ phase diagram of the spin-flop transition in antiferromagnets.

\begin{thebibliography}{99}
\bibitem{comment} The present heat capacity measurement using the thermal relaxation method cannot accurately evaluate the latent heat associated with the first-order transition. Therefore, there remains certain ambiguity in the estimated $S_{\rm N}$.
\end{thebibliography}
\end{document}